\documentclass[prl,12pt,a4paper]{revtex4}
\usepackage[T1]{fontenc}
\usepackage[latin2]{inputenc}
\usepackage{graphicx}
\usepackage{amsmath}
\usepackage{amsfonts} 

\def\ds{\displaystyle}

\begin{document}

\markboth{M. Miśkiewicz, A. Góźdź and J. Dudek}
         {Quantum Rotational Spectra and Classical Rotors}

%
%

\title{QUANTUM ROTATIONAL SPECTRA AND CLASSICAL ROTORS}

\author{
\footnotesize Marek MIŚKIEWICZ\footnote{marekm@kft.umcs.lublin.pl}, 
\footnotesize Andrzej GÓŹDŹ\footnote{gozdz@neuron.umcs.lublin.pl}}
\affiliation{Department of Theoretical Physics, 
University of Maria Curie -- Skłodowska \\ Lublin, Poland}

\author{Jerzy DUDEK\footnote{Jerzy.Dudek@ires.in2p3.fr}}
\affiliation{Institut de Recherches Subatomiques, IN2P3-CNRS and Universit\'e 
Louis Pasteur, Strasbourg\\
France}



\begin{abstract}
We consider the generalized rotor Hamiltonians capable of describing quantum
systems invariant with respect to symmetry point-groups that go beyond the 
usual $D_2$-symmetry of a tri-axial rotor. We discuss the canonical
de-quantisation procedure to obtain the classical analogs of the original
quantum Hamiltonians. Classical and quantum solutions to the Hamiltonians
relevant in the nuclear physics applications are illustrated and compared
using the 'usual' ($D_2$) and an 'exotic' ($T_d$) symmetries.
\end{abstract}
\maketitle
\section{General Aspects}

In this paper we consider the so-called {\em generalized rotor hamiltonians},
i.e. hamiltonians capable of describing the rotating quantum objects that are
invariant under an arbitrary classical point group of symmetry. The word {\em
generalized} implies that we go beyond the properties of the standard
$D_2$-symmetric ('tri-axial') rotors. We also study the classical analogs of
the corresponding quantum objects - 'generalised classical rotors' and the
corresponding trajectories - solutions to the related classical Hamilton
equations of motion.

Hamiltonians of this kind are of special interest in the recent nuclear physics
applications not to mention the traditional ones related to the molecular
physics; realistic microscopic calculations predict the possibility that some
excited nuclear states may be characterised by the high tetrahedral and/or
octahedral symmetries, cf. Refs.\cite{DGS02} and also in: N. Schunck and  J.
Dudek, these Proceedings.

The point groups are subgroups of the orthogonal group, $O(3)$; for the latter
one can define operations of the left- and the right-shifts \cite{1}.
Introducing laboratory and intrinsic frames it can be easily shown that the
left-shift operation corresponds to the rotation in the laboratory- and the
right-shift operation to the one in the body-fixed frame, respectively.  One
can demonstrate that the generators of the left-shift should be interpreted as
the angular momentum operators in the laboratory frame, and the generators of
the right-shift as the operators of the angular momentum in the intrinsic
frame. These two notions allow to introduce the definitions of the rotor
hamiltonians in the mathematically consistent, precise manner. Here we will not
go more into any mathematical details leaving the corresponding discussion to 
a forthcoming publication.

The commutation-relations of the rotation generators (i.e. the angular momentum
operators) are defined in such a way that they obey the convention of
Varshalowich et al.\cite{2}. The latter implies in particular the 'plus' sign
on the right-hand sides of the corresponding commutators for the
intrinsic-frame angular-momentum operators (it is sometimes forgotten that the
often cited in this context the 'minus' sign is a matter of a representation -
and of course no final result depends on that).

\section{Generalised Rotors: Quantum and Classical Hamiltonians}

The nuclear collective rotation can be often described using the rotor
hamiltonian of the form
\begin{equation}
       \hat{H}
       =
       \frac{{\hat{I}_x}^{\;2}}{2\,\mathcal{J}_x}
       +
       \frac{{\hat{I}_y}^{\;2}}{2\,\mathcal{J}_y}
       +
       \frac{{\hat{I}_z}^{\;2}}{2\,\mathcal{J}_z}
                                                                  \label{eqn01}
\end{equation}
where $\hat{I}_{x}$, $\hat{I}_{y}$ and $\hat{I}_{z}$ denote the angular 
momentum operators in the body-fixed reference frame and the three constants,
$\mathcal{J}_x$, $\mathcal{J}_y$ and $\mathcal{J}_z$ are interpreted as the
effective moments of inertia. Such a hamiltonian is invariant with resect to
the $D_2$ point group of symmetry and cannot be used to describe system of 
higher, e.g. tetrahedral symmetry.

In order to provide mathematical means needed to describe the motion of the
higher-symmetry quantum objects we are going to introduce a basis of the 
spherical-tensor operators defined by
\begin{equation}
      \hat{T}^\lambda_\mu \;=\;
      \Big( \big(\ldots \big( \big( 
      \hat{\vec{I}} \otimes \hat{\vec{I}}\; \big)^{\lambda_2=2}
      \otimes \hat{\vec{I}}\; \big)^{\lambda_3=3}
      \otimes \ldots \otimes \hat{\vec{I}} \;\big)^{\lambda_{n-1}=n-1} \otimes 
      \hat{\vec{I}} \; \Big)^{\lambda_n=\lambda=n}_\mu\;,
                                                                  \label{eqn02}
\end{equation}								  			  
with the help of the usual Clebsch-Gordan coupling and the corresponding
generalized rotor hamiltonian as an expansion within the above basis:
\begin{equation}
       \hat{H}
       =
        h_{00} \, T^0_0(2)+
       \sum_{\lambda=1}^\infty 
       \Big( 
            h_{\lambda0} \hat{T}^{\lambda}_0
            \;+\;
            \sum_{\mu=1}^\lambda 
            \big( 
                 h_{\lambda\mu} \hat{T}^\lambda_\mu
                 +
                 (-1)^\mu h^\star_{\lambda\mu} \hat{T}^\lambda_{-\mu}
            \big) 
       \Big)\,,
                                                                  \label{eqn03}
\end{equation}
where $T^0_0(2)=\big( \hat{\vec{I}} \otimes \hat{\vec{I}}\; \big)^0_0$.
Here, the symbol $\{ h_{\lambda\mu} \}$ denotes the full set of the model 
parameters (constants); they however, may be some functions of the scalar 
expressions constructed out of $\vec{I}$. These parameters are
defined in such a way that the resulting hamiltonian is hermitian. Within such
a basis any rotor hamiltonian of any predefined point-group of symmetry can be
easily constructed. Of course the standard rotor expression, Eq.~(\ref{eqn01}),
is a very particular case of that in Eq.~(\ref{eqn03}), see also below.

To obtain a classical description of rotation that is compatible with that 
provided by a given quantum rotor hamiltonian we 'de-quantize' the latter using
the {\em inverse} canonical quantization procedure: 
\begin{equation}
       \frac{1}{i\hbar} [\hat{f},\hat{g}] 
       \longrightarrow
       \{f,g\}\,,
                                                                  \label{eqn04}
\end{equation}
where the symbols $[\,,]$ and $\{\,,\}$ refer to commutators and Poisson
brackets, respectively. The three Euler-angles (formally: operators): 
$\hat{\alpha}$, $\hat{\beta}$, $\hat{\gamma}$ and the corresponding
derivatives: $-i\frac{\partial}{\partial\alpha}$,
$-i\frac{\partial}{\partial\beta}$ and $-i\frac{\partial}{\partial\gamma}$ 
should be replaced by the canonically conjugated variables $\alpha$, $\beta$ and
$\gamma$, and the related momenta $\ell_\alpha$, $\ell_\beta$, $\ell_\gamma$,
respectively.

Having de-quantized angular momenta we obtain a classical Hamiltonian - the
classical analog of the original quantum Hamiltonian that was invariant with
respect to a certain point-group of symmetry - and we can write down the
Hamilton  equations. In principle they form a system of six nonlinear
differential equations. Using the fact that the angular momentum of an isolated
object is conserved in the laboratory frame and selecting the
$\mathcal{O}_z$-axis of the latter to coincide with the direction of the
corresponding angular-momentum vector, the first three of these equations can
be reduced to the following form
\begin{equation}
       \dot{\alpha} = \frac{\partial H}{\partial \ell_\alpha}
                    = A(\beta,\gamma, \vert\vec{J}\vert)\,, 
       \quad
       \dot{\beta}  = \frac{\partial H}{\partial \ell_\beta}
                    = B(\beta,\gamma, \vert\vec{J}\vert)\,,  
       \quad
       \dot{\gamma} = \frac{\partial H}{\partial \ell_\gamma}
                    = G(\beta,\gamma, \vert\vec{J}\vert)\,, 
                                                                  \label{eqn05}
\end{equation}
valid in the body fixed coordinate frame. (Recall: here as always the Euler
angles denote the relative angular position of the body-fixed {\em vs.}
laboratory reference frame, and $\vert\vec{J}\vert$ is the length of the
angular momentum vector - a constant of the motion for a free rotor. Symbols
$A$, $B$ and $G$ denote known functions of their arguments; their explicit
form can only be defined after the symmetry group of interest has been chosen).

As one can see, the above equations do not depend on $\alpha$. Consequently,
solving the system of the second and third of the above equations for 
$\beta=\beta(t)$ and $\gamma=\gamma(t)$ and inserting the solutions into the
equation containing  $\dot{\alpha}$ allows to integrate directly the latter and
obtain the full description of the relative motion of the two reference frames.

Concerning the other three Hamilton equations i.e. those containig
$\dot{\ell}_{\alpha}$, $\dot{\ell}_{\beta}$ and $\dot{\ell}_{\gamma}$: by using
the known relation between the laboratory and the body-fixed reference frames
given by the solutions $\alpha(t)$, $\beta(t)$ and $\gamma(t)$ the components
of $\vec{J}$ in the {\em body-fixed} reference frame can be expressed using
only the length of the angular momentum vector (constant of the motion) and the
angles  $\beta$ and $\gamma$. In other words: because of the conservation of
the angular momentum in the laboratory reference frame, the motion of the
angular  momentum vector in the body-fixed reference frame can be obtained
without  effectively solving the latter three differential equations. Thus
having found solutions only for ${\beta}$ and ${\gamma}$ one obtains the full
description of the motion for the angular momentum vector in the body-fixed
frame.  

In the following we are going to be interested in particular in the equilibrium
points (stable and unstable) associated with the rotational motion; these can
be analyzed within the well established mathematical approach, cf.
e.g. Ref.\cite{Hub90}. The behavior of the rotational motion in the vicinity of the
equilibrium points ($\dot{\beta}=0$, $\dot{\gamma}=0$) allows to classify the
trajectories of the classical rotor and relate them to the quantum spectrum.
These relations will be described qualitatively in the rest of the paper. 

The equilibrium  points can be found from the conditions: $\dot{\beta}=0$,
$\dot{\gamma}=0$.  

In the following we are going to discuss the properties of the $D_2$-symmetric
rotor as a reference object to be compared with the higer symmetry (here:
$T_d$) rotor.

\section{$D_2$-Symmetric Rotor}

First let us consider the second order rotor 
with $D_2$ symmetry: its hamiltonian, Eq.~(\ref{eqn01}), rewritten using the
spherical tensor basis introduced in Eq.~(\ref{eqn02}), takes the form
\begin{equation}
      H_{\textrm{rot}} 
      \,=\,
      h_{00} \, T^0_0(2) +  h_{20} \, T^2_0 +
      h_{22} \, \big( T^2_2 + T^2_{-2} \big),
                                                                  \label{eqn06}
\end{equation}
where $T^0_0(2)=(\vec{J}\otimes\vec{J})^0_0$; there is a one-to-one 
correspondence between the set of the three moments of inertia and the 
$h$-parameters in Eq.~(\ref{eqn06}). In this case the Hamilton equations 
(\ref{eqn05}) read:
\begin{eqnarray}
   	\dot{\alpha} 
        & \;=\; &
	-\frac{J}{3} \, \Big( \, \sqrt{6} \, h_{20} \,-\, 6 h_{22} \,
	\cos (\,2 \gamma \,) \, \Big)\;;
                                                                  \label{eqn07}
                                                                  \\
   	\dot{\beta} 
        &  \;=\; &
	-2 \, J \, h_{22} \, \sin(\beta) \sin(2 \gamma)\;;
                                                                  \label{eqn08}
                                                                  \\
  	\dot{\gamma} 
        & \;=\; &
	J \, \cos (\beta) \, \Big(
	\sqrt{6} \, h_{20} \,-\, 2\,h_{22} \, \cos(2\gamma) \Big)\,.
                                                                  \label{eqn09}
\end{eqnarray} 

One can demonstrate, cf. Ref.\cite{Hub90}, that the motion of the 
rotor in the neighborhood of equilibrium points found through ($\dot{\beta}=0$, 
$\dot{\gamma}=0$), say ($\beta_0,\gamma_0$),
is determined by the eigenvalues of the Jacobi matrix, $\mathcal{M}$, defined 
as follows: suppose that the equations in question have the form
\begin{eqnarray}
     \dot{\beta} &=& B(\beta,\gamma, \vert\vec{J}\vert)\,, \\        
                                                                  \label{eqn10}
     \dot{\gamma}&=& G(\beta,\gamma, \vert\vec{J}\vert)\,.
                                                                  \label{eqn11}
\end{eqnarray}
Then the Jacobi matrix at the equilibrium point satisfies by definition
\begin{equation}
      \mathcal M\,=\,    \left[\begin{array}{cc}
\ds\frac{\partial B}{\partial \beta}  \big\vert_{(\beta_0, \gamma_0)}\,, &
\ds\frac{\partial B}{\partial \gamma} \big\vert_{(\beta_0, \gamma_0)} \\[0.4truecm]
\ds\frac{\partial G}{\partial \beta}  \big\vert_{(\beta_0, \gamma_0)}\,, &
\ds\frac{\partial G}{\partial \gamma} \big\vert_{(\beta_0, \gamma_0)} \\
        \end{array} \right]\,.
                                                                  \label{eqn12}
\end{equation}
It can be shown that in the discussed situation there are the following  two
possibilities. The solutions describing the motion of the angular momentum
vector are periodic (closed trajectories) corresponding to the imaginary
eigenvalues of the Jacobi matrix, or aperiodic ('hyperbolic') ones,
corresponding to the real eigenvalues. In the latter case the equilibrium
points are called saddles, in the former case - centers.  The exact solutions
for saddles are called separatrix trajectories. 

For the illustration below we have selected the rigid-rotor moments of inertia
corresponding to the quadrupole deformations $\beta_2=0.25$ and $\gamma=30^o$,
i.e. maximum triaxiality. 
After straightforward calculations we obtain the following pairs of equilibrium
points for the discussed case: a. the first pair, 
$\{\beta,\gamma\}=\{\pi/2,0\}$ and $\{\beta,\gamma\}=\{\pi/2,\pi\}$, and, b.
the second pair $\{\beta,\gamma\}=\{\pi/2,\pi/2\}$ and
$\{\beta,\gamma\}=\{\pi/2,3\pi/2\}$.

\begin{figure}[ht]
\begin{center}
\includegraphics[scale=1.5]{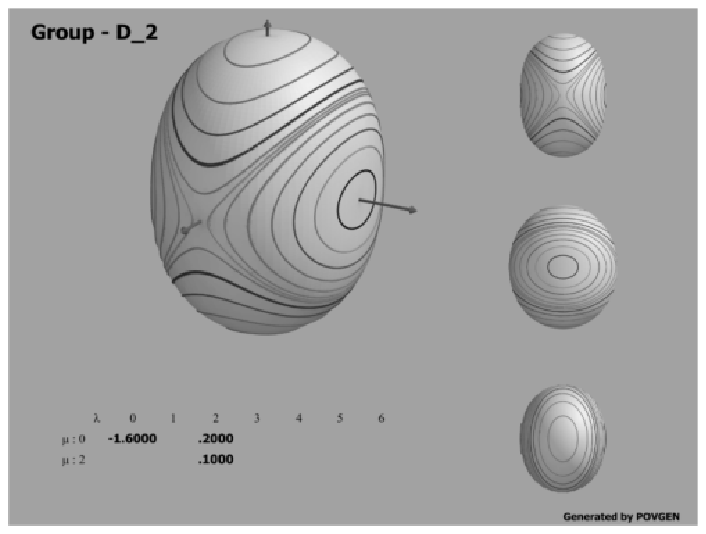} 
\includegraphics[scale=0.612]{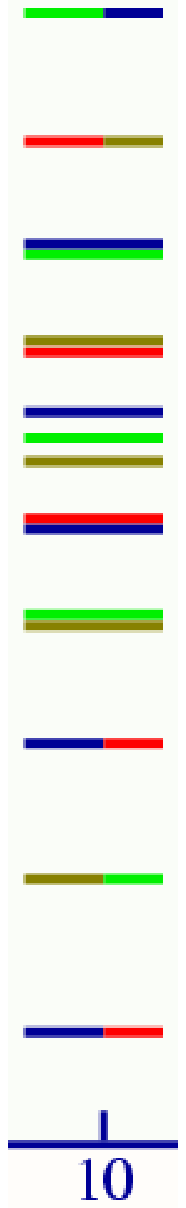} 
\vspace*{8pt}
\caption{
         Energy surface of Eq.~(\ref{eqn12a}) with classical trajectories (left)
         and quantum spectrum (right) for 
         $D_2$-symmetric rotor. Hamiltonian coefficients are: $h_{00}=-0.016$,
         $h_{20}=0.002$, $h_{22}=0.001$, corresponding to the moments of
         inertia $J_x=53.1$, $J_y=67.4$ and $J_z=45.9$ in $\hbar/MeV$ for
         the angular momentum   $J=10\,\hbar$ - i.e. a physical situation
         corresponding to a typical nucleus in Rare Earth nuclei.}
                                                                  \label{fig01}                 
\end{center} 
\end{figure}

The quantum spectra of the corresponding $D_2$-symmetric hamiltonian have the
following very  well known behaviour: the states with the energies approaching
the extremes in the energy scale lie in pairs whose energies are very close to
each other (cf. Fig.~\ref{fig01}). With the energies approaching the middle of
the scale the 'partnership' relation becomes weaker and weaker, corresponding
energies lying further and further apart. On can show that among $(2I+1)$
states in the spin $I$ multiplet, the $2I$ states have a tendency to form the
doublets of partners with very closely lying energies, one state staying always
without a 'partner'. 

The behaviour of the classical rotors can conveniently be discussed with the
help of auxiliary surfaces on which the classical trajectories-solutions can be
drawn. These surfaces are defined as follows. We introduce a body-fixed 
reference frame within which the motion of the angular momentum vector 
$\vec{J}(\alpha,\beta,\gamma)$ can be treated as already known in terms of the
functions $\{ \, \alpha(t)$,   $\beta(t)$, $\gamma(t)\,\}$. To each
orientation of this vector there corresponds rotor's energy 
$E(\alpha,\beta,\gamma)$; the surface composed of all points $\vec\xi$ below
\begin{equation}
      \vec{\xi}
      =
      E(\alpha,\beta,\gamma)
      \frac{\vec{J}(\alpha,\beta,\gamma)}{|\vec{J}(\alpha,\beta,\gamma)|}
                                                                  \label{eqn12a}
\end{equation}
will serve as our reference surface (cf. Figs.~\ref{fig01} and \ref{fig02}).
In other words: this surface is composed of all the points corresponding to
all the solutions of the motion associated to a given length of the vector
$\vec J$; consequently all trajectories - solutions to the Hamilton equations
with $\vert \vec J \vert$-given must lie on such a surface. 

Comparing the quantum spectra for the $D_2$-symmetric Hamiltonian with the
classical solutions whose energies are exactly equal to those of the
corresponding quantum case, one can observe several correlations. First of all,
the energy associated with the separatrix is close to the energy of the
partnerless state and in this sense the two objects can be associated. 
Secondly, states lying above and below the separatrix correspond to the closed
trajectories surrounding centers with maximal and minimal energy. One can
observe higher density of states when energy of the system is approaching the
energy associated with the separatrix.

\section{$T_d$-symmetric Rotor}

The simplest hamiltonian of the generalized rotor with tetrahedral symmetry has 
the form:
\begin{equation}
      \hat{H}_{t} 
      \,=\,
      \underbrace{h_{00} \, T^0_0(2)}_{\ds H_0} 
      +  
      \underbrace{h_{32} \, \big( T^3_2 - T^3_{-2} \big)}_{\ds H_3}\,,
                                                                  \label{eqn13}
\end{equation}
so that to guarantee the hermiticity of the rotor hamiltonian the $h_{32}$
coefficients must be purely imaginary. In the classical limit we have
\begin{equation}
      T^3_2 - T^3_{-2}
      \quad \to \quad
      i\,I_x\,I_y\,I_z\,.
                                                                  \label{eqn13a}
\end{equation}
This structure implies the time-odd character of the latter term since the
$i$-factors disappear and the product of the three $I$-operators changes sign
under time reversal.

The latter feature is worth examining because it implies an interesting
property of the whole hamiltonian in Eq.~(\ref{eqn13}). Since
$[\hat{H}_0,\hat{H}_3]=0$, it follows that the eigenvalues of $\hat{H}_{t}$,
satisfy $E_{t}=E_0+E_3$, where $E_0$ and $E_3$ are the eigenvalues of
$\hat{H}_0$ and $\hat{H}_3$, respectively. We may  write
\begin{equation}
      \hat{H}_{t}\,\Psi_{t} 
      =
      E_{t}\,\Psi_{t}
      \quad \to \quad
      (\hat{H}_0+\hat{H}_3)\,\Psi_{t} 
      =
      E_{t}\,\Psi_{t}       
                                                                  \label{eqn14}
\end{equation}
and at the same time
\begin{equation}
      \hat{\mathcal{T}}\hat{H}_{t}\hat{\mathcal{T}}^{-1}
      =
      \hat{H}_0-\hat{H}_3
      \quad \to \quad
      (\hat{H}_0-\hat{H}_3)\,\Psi_{t}^*
      =
      E_{t}\,\Psi_{t}^*       
                                                                  \label{eqn15}
\end{equation}
and comparing the relations (\ref{eqn14}) and (\ref{eqn15}) we find
\begin{equation}
      \hat{H}_3\,\Psi_{t} = +(E_t-E_0)\,\Psi_{t}
      \quad \leftrightarrow \quad
      \hat{H}_3\,\Psi_{t}^*
      =
      -(E_t-E_0)\,\Psi_{t}^*       
                                                                  \label{eqn16}
\end{equation}
i.e. the spectrum of the $\hat{H}_3$ term in the hamiltonian is symmetric with
respect to zero. Assuming that the energy of the rotational motion is  'purely
cinetic' we must select the paramters of the full hamiltonian  $\hat{H}_t$ in
such a way that the quadtratic (scalar) term dominates so that $E_t$ remains
positive. \\
{\footnotesize In principle the nuclear rotor hamiltonian is  a sum of at
least two terms one of them representing the intrinsic structure  of the
nucleus and the other one the coherent rotational motion; only the sum  of the
two has a priori definite transformation properties with respect to the
symmetry operations (so has the total wave function corresponding to such a
hamiltonian). Following this view point one needs to assume that $h_{32}$
coefficient is a function of group symmetry invariants e.g. $J^2$ and $y^{int}$
the latter describing the internal-coordinate structure of  the rotating body. 
This function is odd under the time reversal operation like the third-order
term in Hamiltonian and the product of them would allow to conserve the time
reversal symmetry. Such an interpretation would modify the approach to the
highly symmeric rotors, here, however, we are not going to discuss this more 
complete model of the nuclear rotor.}
\begin{figure}[ht]
\begin{center}
\includegraphics[scale=1.4]{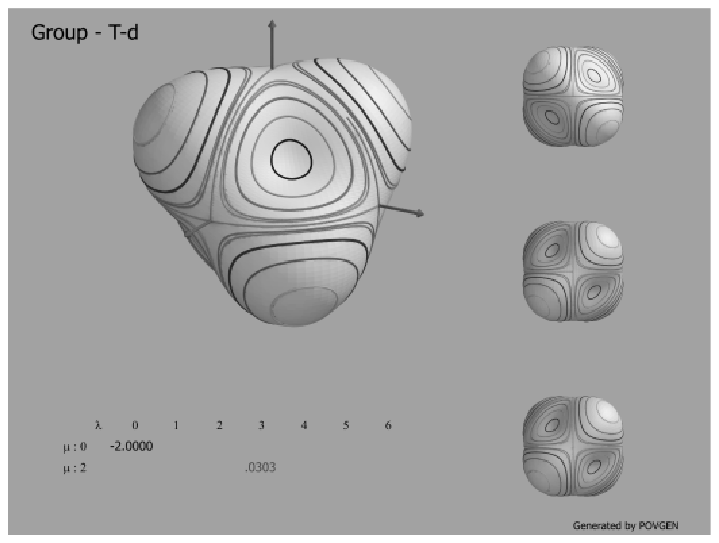} 
\includegraphics[scale=1]{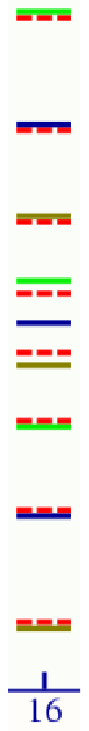} 
\vspace*{8pt}
\caption{
         Energy surface with classical trajectories and quantum spectrum for 
         $T_d$-symmetric rotor. Hamiltonian coefficients are: $h_{00}=-0.016$,
         $h_{32}=i0.003003$, $J=16$.}
                                                                  \label{fig02}                 
\end{center} 
\end{figure} 

The Hamilton equations for the Hamiltonian (\ref{eqn14}) are:
\begin{eqnarray}
   	\dot{\alpha} 
        & \;=\; &
   	-i\, 2\sqrt{3}\, J^2\,h_{32} \,
	    \cos (\beta) \, \sin (2 \gamma)\,;
                                                                  \label{eqn17}
                                                                  \\
   	\dot{\beta} 
        &  \;=\; &
	      -i \,\sqrt{3}\, J^2\, h_{32} \, \cos(2 \beta) \sin(2 \gamma)\,;
                                                                  \label{eqn18}
                                                                  \\
   	\dot{\gamma} 
        & \;=\; &
	   \frac{i\sqrt{3}}{2} \,J^2\,h_{32} \, 
	   \sin(2\gamma)\big(1+3\cos(2\beta)\big)\,;
                                                                  \label{eqn19}
\end{eqnarray}
they are real since the $h_{32}$-coefficient is purely imaginary. 

Calculations show that there are eight equilibrium points for this system
corresponding to all possible combinations of $\beta=\arccos(1/3),
\,\pi-\arccos(1/3)$, $\gamma=\pi/4,\,3\pi/4,\,5\pi/4,\,7\pi/4$ and the four
points related to $\beta=\pi/2$ and $\gamma=0,\,\pi/2,\,\pi,\,3\pi/2$,
classified as saddles.

Analyzing the classical-motion picture, Fig. \ref{fig02}, and the quantum
spectrum one can observe an increase of the density of states near the energy
associated with the separatrix. Another feature of the quantum spectrum,
explained by Eq.~(\ref{eqn17}), is that it is symmetric (in the sense of
Eq.~(\ref{eqn17})) with respect to the state with the energy equal to that of
the separatrix-solution. The symmetry in question corresponds in fact to an
exchange of the roles of the two (out of three) one-dimensional irreducible
representations of the $T_d$-group. In particular one may note that at an even
spin value, I=16 in the illustrated case, all the states below the separatrix
form multiplets with the energy of the three-dimensional irrep {\em above} the
energies of the one-dimensional ones - this tendency is exactly opposite
above the separatrix.

At this stage we have no further comments about
possible analogies between the features of classical  trajectories and e.g.
triple degeneracies of quantum states  corresponding to the
3-dimensional irreducible representation of the  $T_d$-group.

\section{Summary and Conclusions}

We have constructed a mathematical scheme allowing to draw  
certain parallels between the quantum behaviour of high-symmetry rotors and the
solutions of the implied classical equations of the motion for the
angular-momentum vector in the intrinsic (body-fixed) reference frame. The
procedure is analogous to those used in certain studies in molecular physics.
Unlike molecular physics applications where often the semi-classical or
classical limits are of interest ('angular momentum $\to$ $\infty$'),  the
nuclear physics applications must focus on the very low-spin limits, since the
angular-momentum alignment phenomena are expected to destroy the delicate
balance leading to such high symmetries. Moreover, the total symmetries of the
systems are very different in the two types of objects (for instance, the
symmetry with respect to exchange of positions of nuclei/atoms in the molecules
have no nuclear physics analogs). This latter aspect is in general not just a
trivial 'tiny' complication - the full group-structure and thus the related
irreducible representations (even their number), the tensor character of the 
implied wave functions and their symmetry/antisymmetry properties are 
influenced. This can be viewed as a challenge for the nuclear structure
physics whose observables and physics-cases seem to be significantly different
in many respects from those in the the traditional molecular physics 
applications.


\begin{thebibliography}{2}

\bibitem{DGS02}
         J. Dudek, A. G\'o\'zd\'z, N. Schunck and M. Mi\'skiewicz,
         Phys. Rev. Lett. {\bf 88}, 252502 (2002);\\      
         J. Dudek, A. G\'o\'zd\'z and D. Ros\l y,
         Acta Phys. Polonica. {\bf B32} 2625 (2001);\\
         A. G\'o\'zd\'z, J. Dudek and M. Mi\'skiewicz,
         Acta Phys. Polon. {\bf B34} 2123 (2003);\\         
         J. Dudek, A. G\'o\'zd\'z, N. Schunck,
         Acta Phys. Polonica. {\bf B34} 2491 (2003).

\bibitem{1}
A. Barut, R. R\c{a}czka; {\it Theory of Group Representations
and Applications} (PWN, Warszawa, 1977).

\bibitem{2}
D. Varshalovich, A. Moskalev and V. Khershonskii; 
{\it Quantum Theory of Angular Momentum}
(WSP, Singapore, 1988) Ch. 4.2, p.75.

\bibitem{Hub90}
J. Hubbard, B. West; {\it Differential Equations: A Dynamical Systems Approach}
(Springer-Verlag, 1990), Ch. 8.1, p.137.
\end{thebibliography}
\end{document}